\documentclass[10pt,conference,compsoc]{IEEEtran}
\ifCLASSOPTIONcompsoc
  \usepackage[nocompress]{cite}
\else
  \usepackage{cite}
\fi
\ifCLASSINFOpdf
\else
\fi
\usepackage{url}
\usepackage{epsfig}
\usepackage{subfigure}
\usepackage{calc}
\usepackage{amssymb}
\usepackage{amstext}
\usepackage{amsmath}
\usepackage{amsthm}
\usepackage{multicol}
\usepackage{pslatex}
\usepackage{color,soul}

\newenvironment{itemize*}%
  {\begin{itemize}%
    \setlength{\itemsep}{0pt}%
    \setlength{\parskip}{0pt}}%
  {\end{itemize}}

\usepackage[utf8]{inputenc}
\begin{document}
%
\title{Blockchain for Access Control in e-Health Scenarios}

\author{\IEEEauthorblockN{João Pedro Dias}
\IEEEauthorblockA{INESC TEC and\\Faculty of Engineering\\University of Porto\\Porto, Portugal\\
Email:jpmdias@fe.up.pt}
\and
\IEEEauthorblockN{Luís Reis}
\IEEEauthorblockA{INESC TEC\\ FEUP Campus\\ Rua Dr. Roberto Frias\\ Porto, Portugal\\Email:luis.b.pinto@inesctec.pt}
\and
\IEEEauthorblockN{Hugo Sereno Ferreira}
\IEEEauthorblockA{INESC TEC and\\Faculty of Engineering\\University of Porto\\Porto, Portugal\\
Email:hugosf@fe.up.pt}
\and
\IEEEauthorblockN{Ângelo Martins}
\IEEEauthorblockA{INESC TEC\\ FEUP Campus\\ Rua Dr. Roberto Frias\\ Porto, Portugal\\Email:angelo.martins@inesctec.pt}
}


%


\maketitle

\begin{abstract}
Access control is a crucial part of a system's security, restricting what actions users can perform on resources. Therefore, access control is a core component when dealing with e-Health data and resources, discriminating which is available for a certain party. We consider that current systems that attempt to assure the share of policies between facilities are prone to system's and network's faults and do not assure the integrity of policies lifecycle. By approaching this problem with the use of a distributed ledger, namely a consortium blockchain, where the operations are stored as transactions, we ensure that the different facilities have knowledge about all the parties that can act over the e-Health resources while maintaining integrity, auditability, authenticity, and scalability.
\end{abstract}


%
\IEEEpeerreviewmaketitle

\section{\uppercase{Introduction}}
\label{sec:introduction}

\noindent Healthcare is experiencing an explosion of data partially due to the widespread of health data collection systems such as wearables (e.g. fitness trackers)~\cite{lukowicz2004wearable}, health tracking applications (e.g. diet tracking)~\cite{patrick2008health} and ambient assisted living systems such as CAALYX~\cite{boulos2007caalyx}. By now, it is estimated that medical data will grow at a rate of 48\% per year, reaching 2.3 zettabytes by the year of 2020~\cite{emchealth}. 

Lots of new smart objects are empowering the creation of cyber-physical smart pervasive systems with application in multiple domains, including healthcare. These smart objects are under the umbrella of the Internet of Things paradigm, that foresees the advance towards new smart and inter-connected systems by the means of ubiquitous computing~\cite{5579543}.

The explosion of data being collected and, \textit{a posteriori}, analyzed by different entities, leads to the debut of data security and privacy issues. By one hand, these issues are taken into account because such smart devices may be connected to the Internet at some point for accessing its collected data anytime and anywhere~\cite{5579543}. The data being collected from those devices may be part of the Personal Health Records (PHR) and this is typically owned by the patients and may be or not, shared with third-entities ~\cite{tang2006personal}. On the other hand, Electronic Medical Records (EMR) and Electronic Health Records (EHR) store individual information that is required by the healthcare professionals and may be shared among different institutions and facilities~\cite{tang2006personal}.

Hence, there is the need to control the accesses to this data resources by third-entities. Access control is concerned about determining the allowed activities of certain users, mediating every attempt by a user to access a resource in the system~\cite{hu2006assessment}. Dealing with the user access control to health data, personal or medical, holden by different parties, which may be required to be accessed by third-parties with different goals (e.g. insurance companies \textit{versus} doctors), is not an easy task. This is especially problematic when we are still moving towards a unified and interoperable electronic health (e-Health) systems~\cite{7096874}.

In this paper, we suggest an approach to the problem of access control in large scale and distributed systems, as it is observed in e-Health scenarios where different entities and users should be able to access data with different permission levels and granularities. The \texttt{Data Keepers} should be able to manage the accesses to their data by the means of adding, changing or revoking permissions. Such system should be also capable of defining fine-grained permissions both, at the user level and, at the resource level.

The system must also be fault tolerant, which means that it must not be dependable on a centralized architecture. Upon these considerations: consistency, integrity and authenticity of the operations among nodes should be assured. The system must be also immutable providing an accurate audit trail.

This paper is structured as follows: firstly, it is given an overview over the related work in the scope of permission management in e-Health systems, focusing also blockchain approaches for access control. Afterwards is it given a description of the purposed solution architecture. Then we address some core details of the \textit{proof-of-concept} implementation. Finally, some final remarks are presented, summing up the contributions as well as pointing out further developments.

\section{\uppercase{Background \& State of the Art}}

\subsection{\uppercase{Blockchain}}

In our solution, we take into consideration a Distributed Ledger Technology (DLT), specifically, blockchain. A distributed ledger (also known as shared ledger) consists of a consensus of replicated, shared and synchronized digital data distributed along a set of nodes, working as a distributed database, generally geographically dispersed~\cite{Deloitte.2015}. It is important to note that, despite all blockchains being distributed ledgers, not all distributed ledgers are blockchains.

The blockchain is a specific type of distributed ledger conceptualized by Satoshi Nakamoto and used as a core component of the digital currency Bitcoin~\cite{Nakamoto2008}. Data in a blockchain should be tamper-proof, specifically accomplished by the use of cryptography, by the means of digital signatures and digital fingerprints (hashing)~\cite{Deloitte.2015,Nakamoto2008}. Also, consensus must be assured among peers considering scenarios where some of the peers are providing erroneous data, by partially or completed computer/network failures or, even, by malicious intent when some party tries to subvert the ledger~\cite{Nakamoto2008}.

A blockchain consists of a chain of blocks that contains information about transactions. Each one of these transactions, is digitally signed by the entity emitting them. Transactions are combined into a \textit{block}, that is committed to the chain, establishing the blockchain. Each block contains the hash of the previous block, being this propagated along the chain until the first block, created when the blockchain was firstly created, designed \textit{genesis} block~\cite{Nakamoto2008}. 

We can then consider that a blockchain works as a state transaction system (state machine), where there is a state that corresponds to the snapshot of the chain (the result of all transaction until now) and, after adding a new block of transactions to the chain, we got a new snapshot that corresponds to a new state of the system, as result of the new transactions~\cite{buterin2014ethereum}.

In order to a validate a block, it is necessary a \textit{proof-of-work}. This mechanism is used in order to get a consensus in the peer-to-peer network~\cite{Nakamoto2008}. In Bitcoin an HashCash \textit{proof-of-work} is used, being the work effort called \textit{mining}. The \textit{mining} consists of finding a \textit{nonce} (by the means of \textit{brute-force}) that satisfies the condition of generating a digest with the required number of leading zeroes. This \textit{proof-of-work} guarantees consensus in a network following the principle that the nodes will always accept the longest available chain~\cite{Nakamoto2008}. This also implies that older blocks - those further back in the blockchain - are more secure than newer ones. 

There are alternatives to \textit{proof-of-work}. In the \textit{proof-of-stake}, as it is being considered to be used in Ethereum~\cite{buterin2014ethereum}, the creator of the next block to be pushed in the chain is chosen in a deterministic way based on the wealth of the node~\cite{Deloitte.2015}. Another one, as used in the Sawtooth Lake~\cite{TableofC27:online}, uses a \textit{Proof of Elapsed Time} (PoET), which is a lottery-based consensus protocol that takes advantage of the trusted execution environments provided by Intel's Software Guard Extensions.

Notwithstanding the common use of blockchain for trade currencies, like Bitcoin, there exists an array of other applications for the technology. This is possible because, as blockchain is used to store \textit{coin} transactions but it can be used to store any other domain transactions. Furthermore, it can be used as a general-purpose database distributed system, therefore make it useful in a large variety of situations~\cite{Underwood:2016:BBB:3013530.2994581}.

Lastly, blockchains can be considered of three main kinds, as stated by Buterin~\cite{OnPublic5:online}, namely: public, fully-private and consortium. Public blockchains (e.g. Bitcoin), is a type of blockchain in which anyone can read, send transactions to and expect to see them included if they are valid, and, further, anyone in the world can participate in the consensus process. Fully-private blockchains consists of blockchains where write permissions are kept centralized to one organization (even if spread among facilities), existing a closed group of known participants (e.g. a supply chain)~\cite{Underwood:2016:BBB:3013530.2994581}. Finally, consortium blockchains, are partly private in such way that the consensus process is controlled by a number of pre-selected set of nodes. In this type of blockchain, the right to query the blockchain may be public, or restricted to the participants (e.g. governmental institutions and partners). 

\subsection{\uppercase{Access Control}}

The problem of access control has already been covered in the literature. We can observe different ways of controlling and managing accesses in different situations in our everyday technological systems. There exists a problem of defining permission rules, typically known as \textit{policies}, alongside with the problems related with inconsistency, especially in e-Health systems~\cite{Blockchaonline}.

One of the more common approaches is the use of Access Control Lists (ACL), commonly used in modern operating systems. ACL consists of a list associated with an object that specifies all the subjects that can access it, along with the access level (or rights)~\cite{hu2006assessment}. 

Other systems use Access Control Matrix, in which, each row represents a subject, each column an object and each entry is the set of access rights for that subject to that object~\cite{hu2006assessment}.

Specifically in healthcare, Role-Based Access Control (RBAC) and Attribute-Based Access Control (ABAC) have been applied~\cite{Li2010}. RBAC defines the user’s access right basing itself on his/her roles and the role-specific privileges associated with them. The ABAC system extends the RBAC role-based features to attributes, such as properties of the resource, entities, and the environment~\cite{hu2013guide}. Policies in ABAC can be expressed resorting to the eXtensible Access Control Markup Language (XACML), defined by the OASIS consortium~\cite{godik2002oasis}. The XACML standard also includes a reference architecture for designing and implement access control systems, defining the system components and usage flow, that can be used in multiple application domains.

Another approach to access control is the Entity-Based Access Control (EBAC)~\cite{Bogaerts:2015:EAC:2818000.2818009}, which allows the definition of more expressive access control policies. This is accomplished by supporting both, the comparison of attribute values, as well as traversing relationships along arbitrary entities. Moreover, Bogaerts et al. presents \textit{Auctoritas} as an authorization system that specifies a practical policy language and evaluation engine for the EBAC system~\cite{Bogaerts:2015:EAC:2818000.2818009}.

\subsection{\uppercase{Blockchain Applied to Access Control}}

With the widespread use of DLT, and, more specifically, blockchain, novel approaches to resolve the access control problematic appeared using this paradigm, including in e-Health scenarios. Maesa et al.~\cite{Di} proposes a blockchain-based access control, implementing ABAC on top of the blockchain technology, following the XACML reference architecture. This approach validates itself through a reference implementation on top of Bitcoin. However this solution does not encompass the particularities of using such in e-Health context, namely, the possibility of having different authorities and/or entities as resource owners.

In the application of blockchain for e-Health, Yue et al.~\cite{Yue2016} proposes the use of a \textit{Healthcare Data Gateway} (HGD) to enable the patient to own, control and share their data while maintaining data privacy. This solution also encompasses that all the patient e-Health record is stored in a blockchain. Although the novelty of such approach, it implies a disruptive change on the already-existent systems of storing and retrieving e-Health data, what would require a considerable effort to implement which may call into question its current applicability. Also, there are cases where it is needed to access data without the explicit agreement from the patient itself (e.g. due to the patient inability to allow the access or by some governmental requirements) and this solution does not provide the ability to do such (e.g. some family member allow the data access). Also, considering the growth of e-Health data, storing this data on the blockchain itself will result in a rapid growth on its size, exceeding publicly available hard drive capacity, requiring special hardware to full nodes and, further, could lead to the centralization of the blockchain~\cite{chepurnoyspace}.

\section{\uppercase{Illustrative Scenario}}

As of today, there exists an increment of the e-Health data being produced at different sources. Such data can come from more traditional origins like medical exams or medical reports, but, with the advent of the Internet-of-Things, \textit{things} like wearables (e.g. fitness trackers) and ambient assisted living systems, even more data is being produced and consumed by the individual and/or by 3\textsuperscript{rd}-parties. However, it is hard to keep track on the localization of such data as well as when and who is consuming that data. As such, there must exist an access control system transversal to the e-Health domain that allows one to keep data ownership, managing easily the access policies in place. Such system must be clear to the user, as it can easily authorize, deny or revoke permissions on-the-fly (e.g. by the means of a mobile application and push notifications).

\begin{figure}[h]
  \centering
  \includegraphics[width=0.45\textwidth]{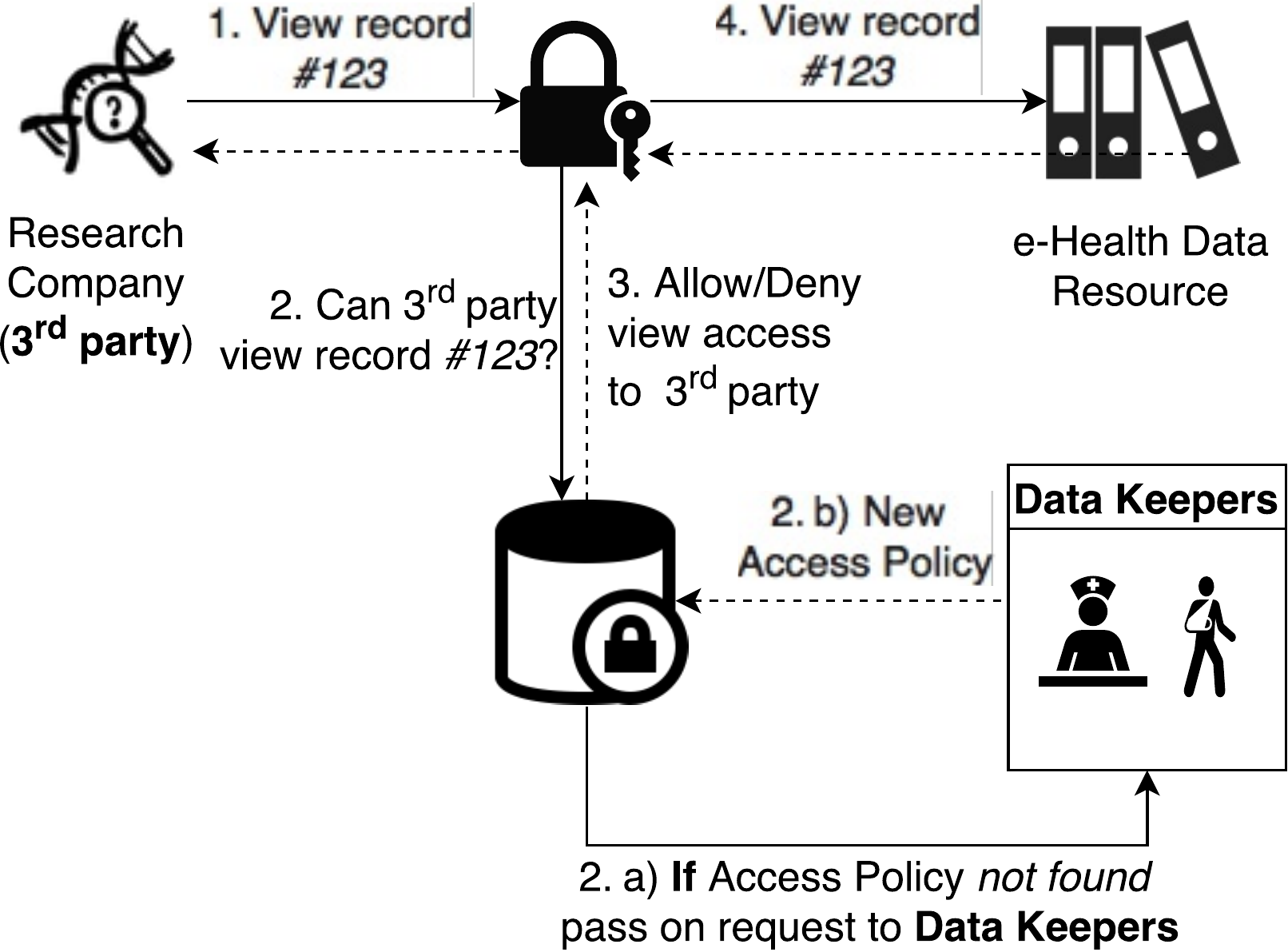}
  \caption{An example of an access to e-Health data, including its different participants. In detail, a third-party entity can request access to a given document, and, in order to do so, the access must be allowed by a request handler system. Such handler should then verify the request by checking an access control policies repository and, further, it must grant or deny the access. If some previously unknown request appears, \texttt{Data Keepers} must issue a new access control policy.}
  \label{fig:case-study}
\end{figure}

An illustrative scenario for observing how an access control system is required to work and, how it would impact the e-Health workflow a case study was designed. A draft of such scene is given on Figure \ref{fig:case-study}. Here we can define the two major actors of such scenario, \textbf{Data Keepwers} and \textbf{Third-Parties}.

\textbf{Data Keepers} is the set of individuals or entities that have ownership over a certain data-entry. They can be, but are not limited to, the data creators, institution responsible, or the data subject(s). The possibility of defining different sets of keepers is a requirement since, depending on the e-Health record or source, different restrictions may apply. Namely, when dealing with PHR data the only keeper is the individual or the legal responsible~\cite{nahitreport}. However, EMR data is controlled by a set of clinicians and staff within one health care organization, and, further, EHR data can even be shared between more than one health organization~\cite{nahitreport}.

\textbf{Third-parties} are any organization, entity or individual that have interest in accessing the data (e.g. insurance companies or research institutes). Generally, these consumers have time-limited access to a small portion of data of a dataset or to a specific individual.

In detail, we can sum up the interaction of the different parties and the system as it follows (using the Figure \ref{fig:case-study} as reference):
\begin{enumerate}
\item{There is a \texttt{3\textsuperscript{rd} Party} who have interest in viewing a specific e-Health record (in Figure \ref{fig:case-study} as example we use the document id \textit{\#123}). This party asks to an intermediate if it can access the record mentioned;}
\item{The intermediate checks the access against an Access Policy repository where it matches with already existent policies. If there is no Access Policy within a particular third-entity and a document, an extra two steps occurs: (1) A notification with the details of the access request is sent to the set of document keepers;
(2) The access keepers must allow or deny the access request. If there is more than one keeper on the document, a consensus must be reached. This consensus depends on the proprieties of the document and it can require the approval of one keeper, majority or all.}
\item{After checking the request against the Access Policies repository, the intermediate grants or denies access to the document by the 3\textsuperscript{rd}-party;}
\item{If access is allowed, the \texttt{3\textsuperscript{rd} Party} can now view the specific e-Health record (in Figure \ref{fig:case-study} the document id \textit{\#123}). If not, the request is denied and the entity cannot access the document in any way.}

\end{enumerate}

Taking into account the interaction flow we can say that the scenario shares the three main components of common access control systems (specified in XACML standard), namely, a Policy Decision Point (PDP), a Policy Administration Point (PAP) and a Policy Enforcement Point (PEP)~\cite{godik2002oasis}. 

Here, the PEP is the intermediate negotiator that intercepts the users' requests and enforces the PDP's decision. Further, the PDP is responsible for evaluating the requests against an Access Policies repository. However, in our case, there is not a traditional PAP, because there is no central authority managing the policies, being each one of the document keepers managers of their own documents, working as a distributed PAP system.

In this paper, we leverage the use of blockchain as a repository for Access Policies, and, furthermore, as a way of enabling the existence of a distributed PAP component.




\section{\uppercase{Proposed Approach}}

Our approach consists of using Blockchain technology as a way to accomplish a more reliable and user-empowered solution for access control management in an e-Health environment. Such approach allows us to define fine-grained access control while maintaining the consensus in a distributed system, authenticity, immutability and auditability. 

A \textit{proof-of-concept} of the approach hereby described and detailed was implemented in order to verify its feasibility.

In our solution, we use an approach similar to the Access Control Matrix, which allows the establishment of a correspondence between a subject, an object and a set of rights. However, this information is not stored as is, due to the inherent proprieties of the use of blockchain. As a transaction-based state machine, we store transactions corresponding to a pre-defined set of the state machine on the Access Policies repository. 

\subsection{Access Control Model} \label{accessmodel}

\begin{figure}[h]
  \centering
  \includegraphics[width=0.48\textwidth]{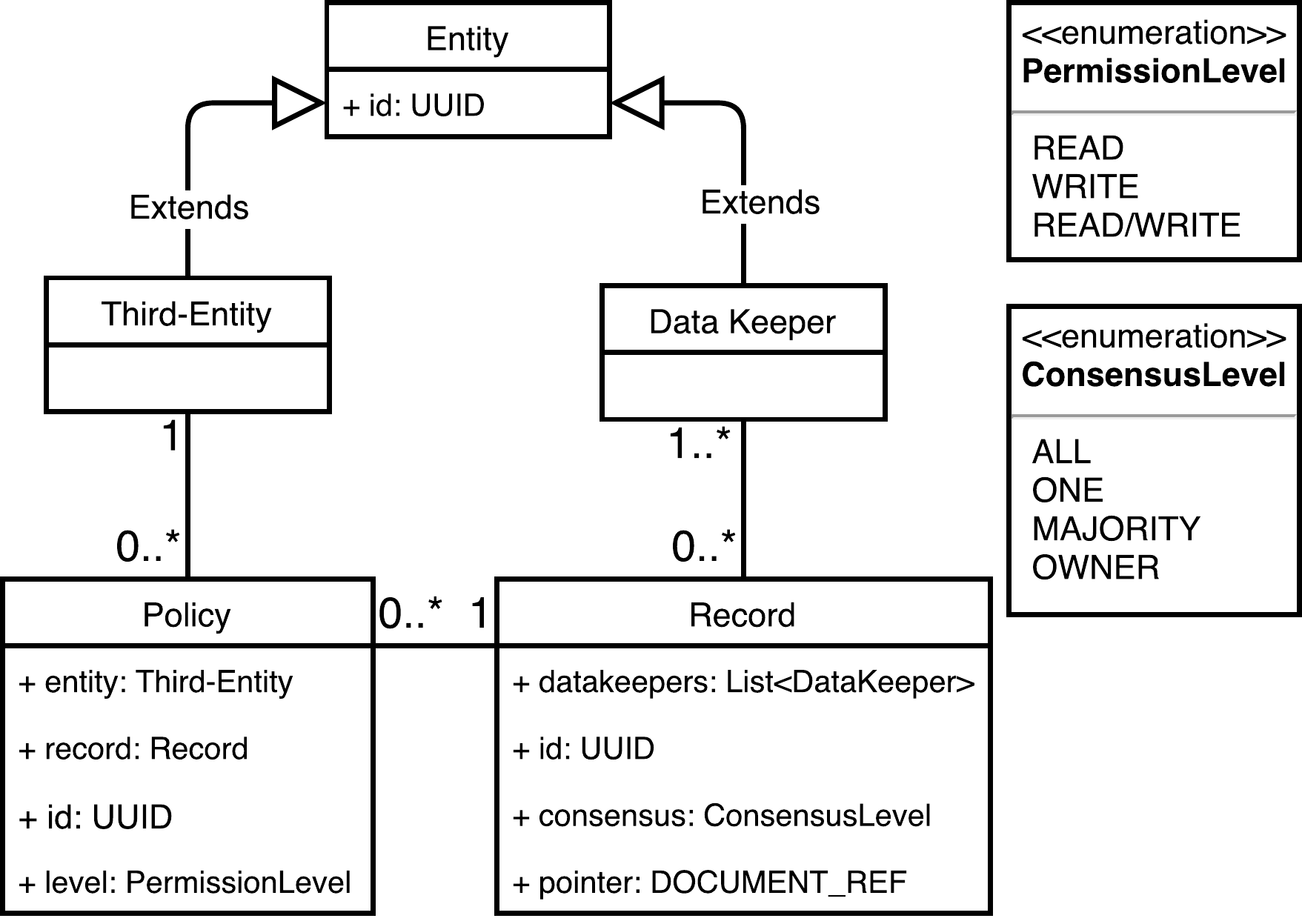}
  \caption{UML diagram specifying the classes of the system and their relationships.}
  \label{fig:full-uml}
\end{figure}

Upfront for defining our model, it is needed to define all the entities and relationships enrolled in it. Such model is given in Figure \ref{fig:full-uml}, and five classes can be identified in our approach. 

First off, we have \textit{Entities} that can be \texttt{3\textsuperscript{rd} Party}'s or \texttt{Data Keepers}. Further, we have \texttt{Policies} and \texttt{Records}. Each \texttt{Policy} refers to a relation between one and only one \texttt{3\textsuperscript{rd} Party} and an e-Health \texttt{Record}, with the respective level of access, \texttt{PermissionLevel}. In turn, each \texttt{Record} can have one or more \texttt{Data Keepers} that have partial or total ownership over it. 

\begin{figure}[h]
  \centering
  \includegraphics[width=0.4\textwidth]{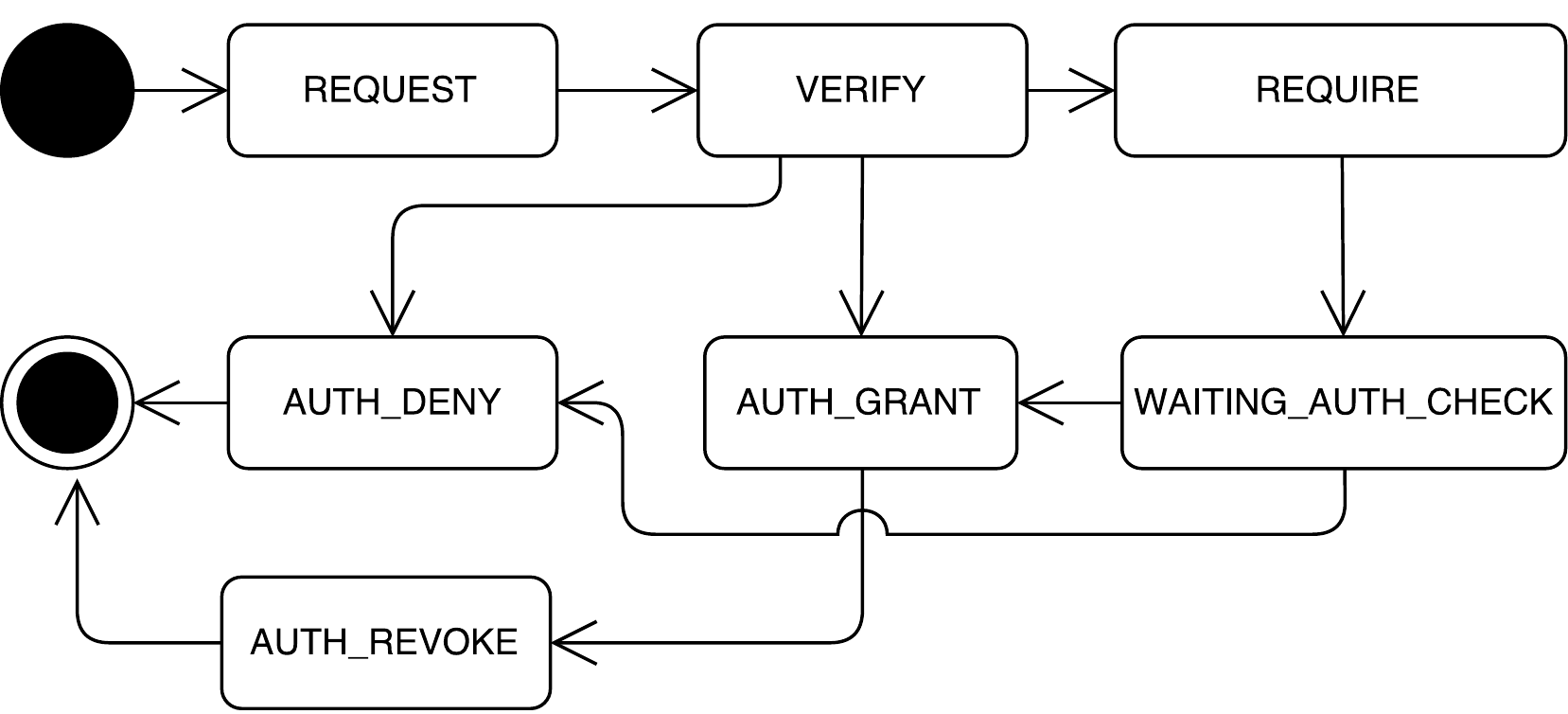}
  \caption{State machine diagram representing the lifecycle of an access request by a \texttt{3\textsuperscript{rd} Party}. The composite state \texttt{AUTH\_CHECK} represents the individual authorization requests needed by the \texttt{Data Keepers} of the record being queried.}
  \label{fig:full-sm}
\end{figure}


\begin{figure}[h]
  \centering
  \includegraphics[width=0.35\textwidth]{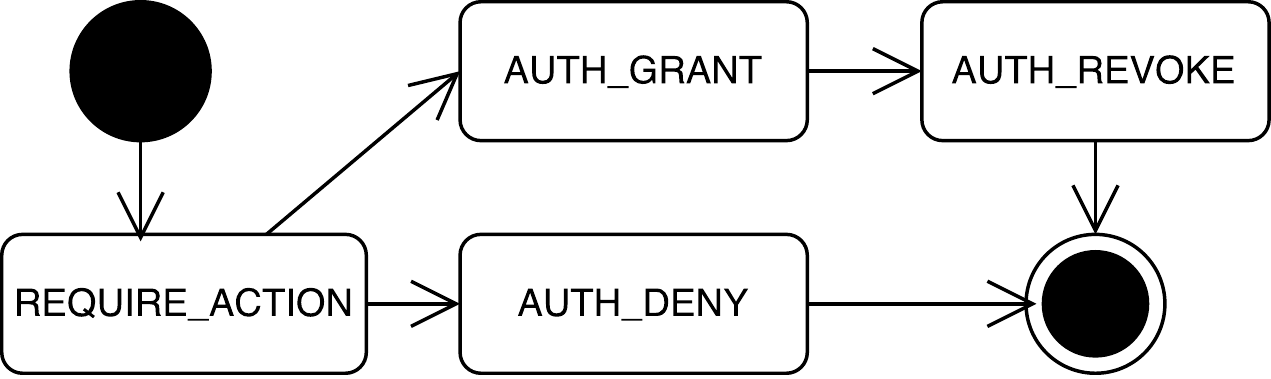}
  \caption{State machine diagram representing the individual authorization lifecycle.}
  \label{fig:indv-sm}
\end{figure}

Moreover, the information related to the model is preserved by the means of storing transactions (since this model is compatible with the use of blockchain). The transactions contain information about with 3 different state machines, that have dependencies between them, being always related to the class model defined.

\textbf{Access Policy State Machine}, as represented in Figure \ref{fig:full-sm}, is the state machine related to the main logic of creating Access Policies.

\textbf{Record Life-cycle State Machine} is responsible for the operations over an e-Health record. The record lifecycle begins with its creation, \texttt{CREATE}, then it can suffer diverse updates, \texttt{UPDATE}, until it is removed, \texttt{REMOVE}.

\textbf{Individual Authorization State Machine}, as represented in Figure \ref{fig:indv-sm}, describes the lifecycle for each user access grants over a given \texttt{REQUIRE}, which lead to a number of instantiations equal to the number of \texttt{Data Keepers} required. Each individual instantiation evokes a \texttt{REQUIRE\_ACTION}, then the \texttt{Data Keeper} can allow (\texttt{AUTH\_GRANT}) or deny (\texttt{AUTH\_DENY}) the access. Eventually, before reaching the final state, the \texttt{Data Keeper} can revoke (\texttt{AUTH\_REVOKE}) a previously granted authorization.

As stated, the main logic of attributing Access Policies for \texttt{3\textsuperscript{rd} Party} access an e-Health record is controlled by the \textit{Access Policy State Machine}, detailed in Figure \ref{fig:full-sm}. This state machine jumps the \texttt{init} state when an access request, \texttt{REQUEST} from a \texttt{3\textsuperscript{rd} Party} enters the PEP. Then, the \texttt{REQUEST} is verified, \texttt{VERIFY}, against the already existent information on the blockchain (by the means of a snapshot operation). If, and only if, the information about this particular access is present in the snapshot, the request can be granted, \texttt{AUTH\_GRANT} or denied \texttt{AUTH\_DENY}.  

Additionally, if there is no information about an access request, the Access Policy must be required, \texttt{REQUIRE}, by the means of checking the number of permissions required form the texttt{Data Keepers} in order to get a consensus (specific to the record), using for that the \textit{Individual Authorization State Machine}. This means that there is no central authority authorizing requests from \texttt{3\textsuperscript{rd} Parties}, and it is required that some set of \texttt{Data Keepers} allow the access. While this process is running, the state machine enters into a waiting state, \texttt{WAITING\_AUTH\_CHECK}.

At last, there can be a point in the future when it is needed to revoke a previously granted access, \texttt{AUTH\_REVOKE}. The \texttt{final} state is, by this, reached by the existence of an \texttt{AUTH\_REVOKE} or an \texttt{AUTH\_DENY}.

\subsection{Block Model}

As our Access Control follows the model presented in Subsection \ref{accessmodel}, based upon transactions, we leverage the use of blockchain as a repository for these transactions. This way we never have a current state of the permissions written in the repository, but, notwithstanding, we can take a snapshot to the blockchain at any moment and, as result, check the Access Policies in place.

The basic structure of a block in our chain is the following:
\begin{itemize}
\item \texttt{index}: Corresponds to the index of the current block in the blockchain.
\item \texttt{timestamp}: Timestamp corresponding to when the block was generated.
\item \texttt{previousHash}: Hash of the previous block in the chain.
\item \texttt{digitalSign}: Digital signature of the current block data.
\item \texttt{data}: Content of the block. Corresponds to a set of transactions describing the access control policies, records information and individual authorizations. 
\item \texttt{nonce}: Value that is set so that the hash of the block will contain a run of leading zeros. This value is calculated iteratively until the required number of zero bits in the \texttt{hash} is found. This requires time and resources, making it so that the correct \textit{nonce} value constitutes \textit{proof-of-work}. 
\item \texttt{hash}: A SHA256 hash corresponding to the block data. This hash must have a leading \textit{a priori} defined sequence being this leading sequence what defines the effort of the \textit{proof-of-work}. In Bitcoin, this leading sequence corresponds to a certain number of zeros in the beginning of the hash.
\end{itemize}

Additionally, focusing on our approach, the \texttt{data} field should be detailed, as it is the field that serves as transaction information storage. This \texttt{data} field includes in it three sub-fields, namely: 
\begin{itemize}
\item \texttt{records}: Transaction information relative to transactions of the state machine presented in Figure \ref{fig:doc-sm}, about creation, update or deletion of e-Health records of any kind.
\item \texttt{policies}: Transaction information relative to transactions of the state machine presented in Figure \ref{fig:full-sm}, about creation and revocation of access policies.
\item \texttt{individualAuths}: Transactions about individual authorization by each one of a \texttt{Record} \texttt{Data Keeper} in relation to each \texttt{Policy}. 
\end{itemize}

The use of hashes allows us to maintain integrity along the immutable chain of transactions without a central authority, since any change in the data would result in a different hash, invalidating the next blocks in the blockchain. Additionally, as result, we can also achieve accountability and auditability. Authenticity is assured by the assign of a key-pair to each entity with access to the blockchain, identifying who write each block in the chain.  

\subsection{Architectural Design}

\begin{figure}[h]
  \centering
  \includegraphics[width=0.425\textwidth]{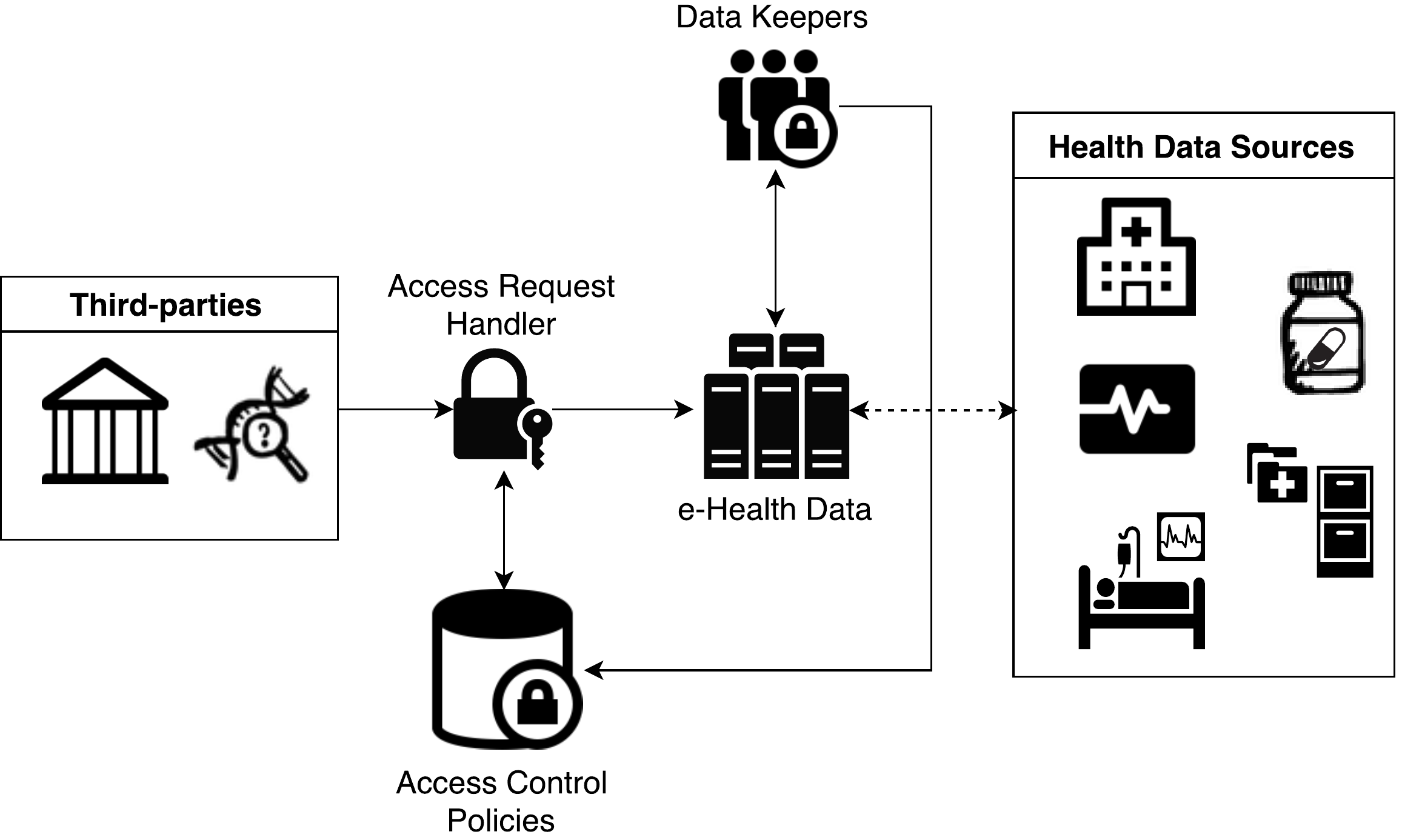}
  \caption{Overview of the system architecture and interactions. The e-Health data access requests treated seamlessly by the handler. Additionally, the resources (e-Health data) are, typically, associated with one facility or repository, but all the available resources are known by all nodes (facilities).}
  \label{fig:summary-arch}
\end{figure}

At the system architectural level, our approach uses blockchain, being distributed by default, working as a peer-to-peer network connecting the different nodes, corresponding to the diverse facilities or organizations that can store, create or/and change e-Health data. Nodes in the network synchronize between them by following a set of rules:
\begin{itemize}
\item When a new block is generated by a node, this block is broadcasted to the network;
\item When a node connects to a new peer in the network it queries for the latest block;
\item If a node finds a block that has a higher index than the last known block, it either adds the block to its current blockchain (in the case of the difference is equal to one node) or queries another node to get the full blockchain.
\end{itemize}

The system implements a consortium blockchain\cite{OnPublic5:online}, which means that the blockchain is partly private, in the way that instead of allowing any person to participate in the process of transaction verification or, allowing only one entity to have full control, a few selected nodes are predetermined, providing the same benefits associated with private blockchain. The \texttt{Data Keepers} and \texttt{3\textsuperscript{rd} Party} can then interact with the system by the means of using public available Application Programming Interfaces (API's) or applications designed to do such.

Furthermore, despite the use of XACML standard for access control systems and blockchain as storage, there are some architectural choices due to the e-Health domain restrictions and more complex use cases. An overview can be observed in Figure \ref{fig:summary-arch}. 

Focusing on the \textit{e-Health Data}, as of today, this data is not aggregated in one storage, being spread by multiple institutions and organizations. As such, every time that any \texttt{3\textsuperscript{rd} Party} requests access to a specific record there is the need of locating this information, and then, proceed to check if the request is already approved or if there is the need to create a new access policy. So, as an improvement, information about the creation of new records must be kept and spread along all the organizations and facilities in such way that a request to a resource can be handled by any member of the private blockchain. 

Additionally, aiming attention to the \texttt{Data Keepers}, it was noticed that there is a set of situations where the ownership of e-Health data records is not explicit to only one entity but shared among more than one entity or individual, as is the case of EMR \textit{versus} PHR. Taking this into account, we set up a mechanism of consensus when creating new access policies. Each e-Health record has a level of agreement that must be achieve before allowing a \texttt{3\textsuperscript{rd} Party} the access to a \texttt{Resource}, being this level associated with the \texttt{Resource} itself (Figure \ref{fig:full-uml}). Then, there is a number of executions of the individual authorization state machine (Figure \ref{fig:indv-sm}) corresponding to the number of that \texttt{Record} \texttt{Data Keepers}. Reaching the minimum number of individual authorization (that can be either \texttt{AUTH\_GRANT} or \texttt{AUTH\_DENY}), the access request state machine (Figure \ref{fig:full-sm}) will create an access policy accordantly with the consensus reached (that can be, once more, either \texttt{AUTH\_GRANT} or \texttt{AUTH\_DENY}).

\begin{figure*}[h]
  \centering
  \includegraphics[width=0.75\textwidth]{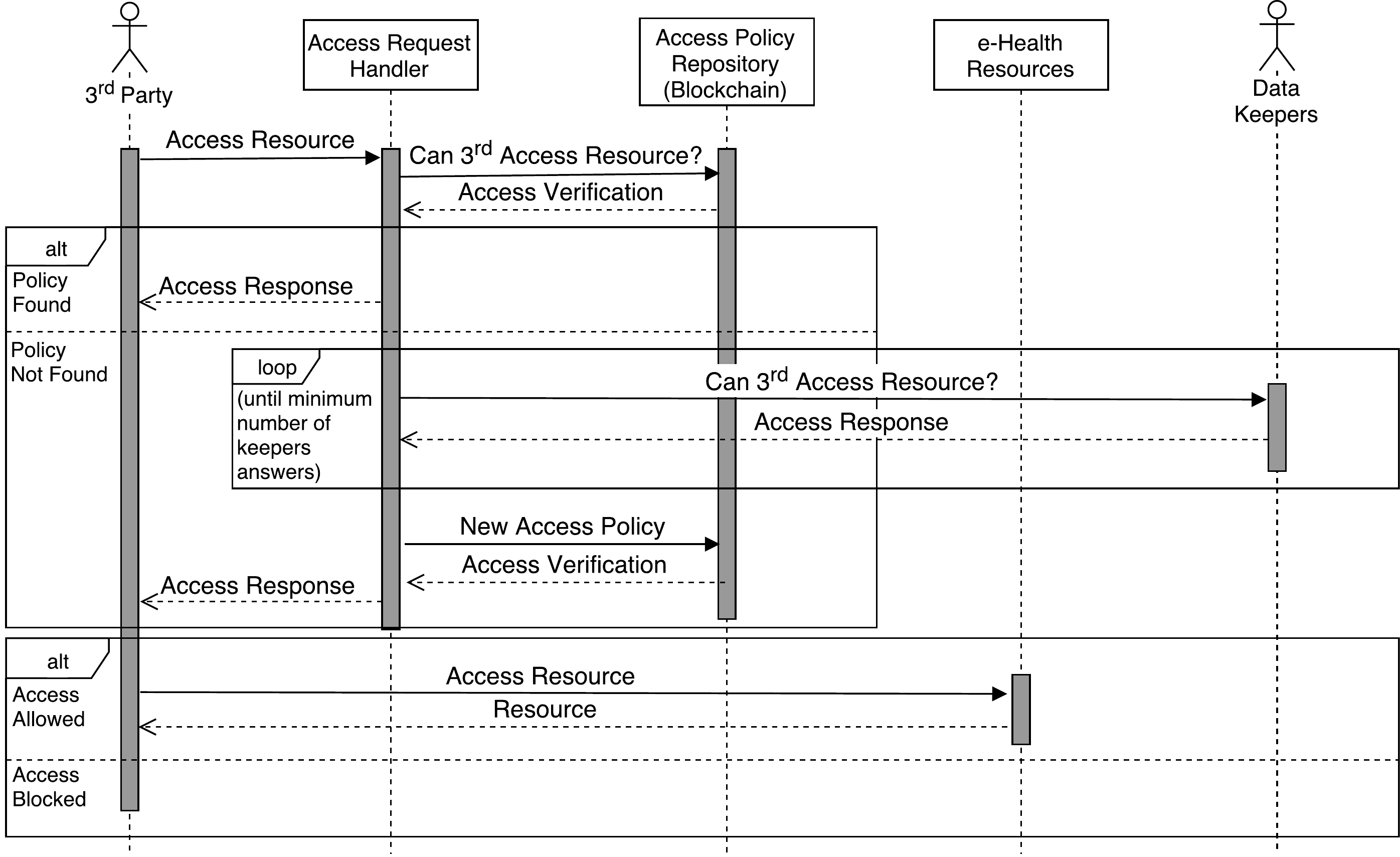}
  \caption{Sequential view on how some \texttt{3\textsuperscript{rd} Party} can access or request access to an e-Health resource, detailing the communication between the inner modules of the architecture. It is also visible the process of creation of new access control policies.}
  \label{fig:sequence}
\end{figure*}

From the functional architecture viewpoint, we can sum up the system interaction as stated in the sequence diagram of the Figure \ref{fig:sequence}. This diagram describes the process of a \texttt{3\textsuperscript{rd} Party} requesting access to an e-Health \texttt{Record}, owned partially or totally by one or more \texttt{Data Keepers}. Further, the diagram describes both the case of checking against an already existent access policy of the \texttt{3\textsuperscript{rd} Party} and the \texttt{Record} or the process of creating a new \texttt{Policy} by checking the necessary number of \texttt{Data Keepers}.

\subsection{Security Threat Analysis}

In our approach, there are a number of security questions that must be taken into consideration since we have to consider that mistakes can be made by, for example, careless operators. Alongside with threats coming from human mistakes, there is the need of considering also faults introduced by intentional system manipulation coming from individuals with malicious intent. Note that we assume that system's operators do not have physical access to the machines where the Access Control system is deployed, communicating only by an existent GUI/CLI.

\paragraph{Why existing Access Control system storage do not suffice?} Currently deployed systems use a centralized or almost-centralized (\textit{quasi}-decentralized) solution to store and access control policies, independently on how these policies are described (e.g. ACL, RBAC or EBAC). Such systems are vulnerable to network or machine failures since in case of failure all the facilities are affected by the impossibility of validating access control policy rules over their resources. The use of a DLT as a purely-distributed database for storing the policies allows the correct functioning of all non-falling facilities and automatically synchronizes when the normal function of the system is restored. Additionally, even if a node is compromised, the threat is confined to the node specifically, not compromising the whole network.

\paragraph{Why do Access Control systems have difficulties ensuring the validity of policies \textit{lifecyle}?} Access Control systems typical record the addition, revoking or change in a policy by the means of logging. Such approach is vulnerable to modifications by a malicious third-party because there is no way of assuring the integrity of these logs. The use of a blockchain as a way of storing policies as transactions, assures us that older policies (and operations over them) have integrity and are frequently recomputed from a set of trusted commands. Additionally, we can ensure that, as time passes (and the blockchain grows), the blocks are more secure as you go further back in the chain.

\paragraph{What does it take to a third-entity with malicious intent to add a rogue policy?} As the access to resources present in the e-Health system must be validated by the means of a policy in the Access Control system, successful violation by a malicious party is unconditionally dependent on the injection of a rogue permission into the system. Such attempt can be mainly made by two attack vectors: (i) by attempting to mine a new block in the blockchain, though our approach makes that unlikely due to the usage of a consortium blockchain, where only allowed entities can add blocks; hence, there is the need of such malicious entity to take over more than 51\% of the entire blockchain network to accomplish such task; and (ii) by compromising an Access Control system machine or any third-application that communicates with it, allowing a party to submit one or more rogue policies to the system. This last type of attack cannot be prevented by the blockchain, and thus relies on good security practices on edge applications and devices.

\paragraph{What are the privacy implications of having a transparent ACL system?} As the data stored in the blockchain is available to all nodes that belong to it, some privacy concerns can arise. However, since the approach consists of a consortium blockchain, all the nodes are, \textit{a priori}, well-known and trustworthy, mitigating the privacy concerns of a public blockchain.

\section{\uppercase{Implementation Details}}

During the implementation of the \textit{proof-of-concept}, some decisions were made; we describe these details here with the intent of helping the reader to re-implement a similar \textit{proof-of-concept} or production-ready system. 

The whole system, from the transactions logic to the writing to the blockchain, was implemented using \texttt{Javascript} due to the simplicity of the language and availability of libraries. Such libraries, as for example, the built-in \texttt{crypto} module provides us with a mechanism to digitally sign the blocks payload using public-key cryptography (using RSA-SHA256) and to calculate the hashes of each block using SHA256 algorithm.

From our perspective and due to the huge disparity among data types, three main classes were created: \texttt{Document}, \texttt{Entity} and \texttt{Transaction}. 

The information of the State Machines mentioned in section~\ref{accessmodel} is given by the \texttt{Transactions} data present in each moment and not by the states of the machine itself. Thus, the current state of the system given by the process of \textit{snapshot}. A \textit{snapshot} is like a picture taken from the blockchain,  consisting of applying all previous transactions until now in order to get the state of all operations. As example, for evaluate a access request requires the execution of all transactions about that specific \texttt{3\textsuperscript{rd} Party}, \texttt{Record} and associated \texttt{Data Keepers}.

The implemented blockchain uses \textit{proof-of-work} that bases itself on a \textit{brute-force} mechanism of hash calculation. This mechanism works as the \textit{nonce} is iteratively incremented until the resulting SHA256 hash matches the \textit{a priori} defined number of leading zeroes --- this is similar to the Bitcoin system and establishes the ``effort''. 

However, we can easily tweak the ``effort'' to better suit our use-case. Although we implemented the \textit{proof-of-work} mechanism, due to its  simplicity, there are alternatives as \textit{Proof of Elapsed Time} (PoET) or \textit{proof-of-stake} mechanisms, that are not as resource consuming as the implemented one, which can be better to validate the transactions in the context of a consortium blockchain applied to e-Health scenarios.

One of the details that required some attention was the writing of \textit{blocks} to the blockchain. As previously mentioned, each transaction has a unique identifier (\texttt{ID}) to allow the identification of the same transaction over its different states. This helps to prevent different states of the same transaction from being written to the same block.


\section{\uppercase{Conclusions}}
\label{sec:conclusion}

\noindent In this paper, we present an approach to solving the problem of managing access control in the e-Health ecosystem. Access Control is a special complex task in e-Health since resources and data are distributed among different facilities and institutions. Further, this is even more problematic because in some cases, e-Health resources are not owned or managed by a single entity or individual. As a way of overcoming this complexity, we propose an approach that leverages the use of blockchain for store transactional information about e-Health records and access control policies.

For purposes of supporting the plausibility of our solution, a \textit{proof-of-concept} was designed and implemented. This \textit{proof-of-concept} allowed us to make some, even if preliminary, tests and validations over our approach from a functional and applicational perspective.

Overall, we determine that our approach is viable, giving diverse advantages when comparing to the in-place systems. This advantages although, not limited to, includes the integrity, transparency, and authenticity of the access control policies in the system, being, this information distributed and synchronized by all the institutions and organizations that make part of the consortium.


\section*{\uppercase{Acknowledgements}}

\noindent This work was supported by Project “NanoSTIMA: Macro-to-Nano Human Sensing: Towards Integrated Multimodal Health Monitoring and Analytics/NORTE-01-0145-FEDER-000016" financed by the North Portugal Regional Operational Programme (NORTE 2020), under the PORTUGAL 2020 Partnership Agreement, and through the European Regional Development Fund (ERDF).

\bibliographystyle{IEEEtran}
\bibliography{example}

\begin{thebibliography}{10}
\providecommand{\url}[1]{#1}
\csname url@samestyle\endcsname
\providecommand{\newblock}{\relax}
\providecommand{\bibinfo}[2]{#2}
\providecommand{\BIBentrySTDinterwordspacing}{\spaceskip=0pt\relax}
\providecommand{\BIBentryALTinterwordstretchfactor}{4}
\providecommand{\BIBentryALTinterwordspacing}{\spaceskip=\fontdimen2\font plus
\BIBentryALTinterwordstretchfactor\fontdimen3\font minus
  \fontdimen4\font\relax}
\providecommand{\BIBforeignlanguage}[2]{{%
\expandafter\ifx\csname l@#1\endcsname\relax
\typeout{** WARNING: IEEEtran.bst: No hyphenation pattern has been}%
\typeout{** loaded for the language `#1'. Using the pattern for}%
\typeout{** the default language instead.}%
\else
\language=\csname l@#1\endcsname
\fi
#2}}
\providecommand{\BIBdecl}{\relax}
\BIBdecl

\bibitem{lukowicz2004wearable}
P.~Lukowicz, T.~Kirstein, and G.~Troster, ``Wearable systems for health care
  applications,'' \emph{Methods of Information in Medicine-Methodik der
  Information in der Medizin}, vol.~43, no.~3, pp. 232--238, 2004.

\bibitem{patrick2008health}
K.~Patrick, W.~G. Griswold, F.~Raab, and S.~S. Intille, ``Health and the mobile
  phone,'' \emph{American journal of preventive medicine}, vol.~35, no.~2, p.
  177, 2008.

\bibitem{boulos2007caalyx}
M.~N.~K. Boulos, A.~Rocha, A.~Martins, M.~E. Vicente, A.~Bolz, R.~Feld,
  I.~Tchoudovski, M.~Braecklein, J.~Nelson, G.~{\'O}. Laighin \emph{et~al.},
  ``Caalyx: a new generation of location-based services in healthcare,''
  \emph{International journal of health geographics}, vol.~6, no.~1, p.~9,
  2007.

\bibitem{emchealth}
IDC, ``The digital universe: Driving data growth in healthcare,'' EMC
  Corporation and International Data Corporation, Report, 2014.

\bibitem{5579543}
L.~Tan and N.~Wang, ``Future internet: The internet of things,'' \emph{2010 3rd
  International Conference on Advanced Computer Theory and
  Engineering(ICACTE)}, vol.~5, pp. V5--376--V5--380, Aug 2010.

\bibitem{tang2006personal}
P.~C. Tang, J.~S. Ash, D.~W. Bates, J.~M. Overhage, and D.~Z. Sands, ``Personal
  health records: definitions, benefits, and strategies for overcoming barriers
  to adoption,'' \emph{Journal of the American Medical Informatics
  Association}, vol.~13, no.~2, pp. 121--126, 2006.

\bibitem{hu2006assessment}
V.~C. Hu, D.~Ferraiolo, and D.~R. Kuhn, \emph{Assessment of access control
  systems}.\hskip 1em plus 0.5em minus 0.4em\relax US Department of Commerce,
  National Institute of Standards and Technology, 2006.

\bibitem{7096874}
W.~O. Nijeweme-d'Hollosy, L.~van Velsen, M.~Huygens, and H.~Hermens,
  ``Requirements for and barriers towards interoperable ehealth technology in
  primary care,'' \emph{IEEE Internet Computing}, vol.~19, no.~4, pp. 10--19,
  July 2015.

\bibitem{Deloitte.2015}
Deloitte., ``{Bitcoin, Blockchain {\&} distributed ledgers: Caught between
  promise and reality},'' Centre for the Edge, Australia, Tech. Rep., 2015.

\bibitem{Nakamoto2008}
S.~Nakamoto, ``{Bitcoin: A Peer-to-Peer Electronic Cash System},'' p.~9, 2008.

\bibitem{buterin2014ethereum}
F.~Ethereum, ``Ethereum: A next-generation smart contract and decentralized
  application platform,'' 2014.

\bibitem{TableofC27:online}
I.~Corporation, ``Sawtooth lake latest documentation,'' 2015, (Accessed on
  06/02/2017).

\bibitem{Underwood:2016:BBB:3013530.2994581}
\BIBentryALTinterwordspacing
S.~Underwood, ``Blockchain beyond bitcoin,'' \emph{Commun. ACM}, vol.~59,
  no.~11, pp. 15--17, Oct. 2016. [Online]. Available:
  \url{http://doi.acm.org/10.1145/2994581}
\BIBentrySTDinterwordspacing

\bibitem{OnPublic5:online}
V.~Buterin, ``On public and private blockchains,'' August 2015, (Accessed on
  06/06/2017).

\bibitem{Blockchaonline}
\BIBentryALTinterwordspacing
R.~Krawiec and M.~White, ``Blockchain: Opportunities for health care,'' August
  2016, (Accessed on 06/05/2017). [Online]. Available:
  \url{https://www2.deloitte.com/us/en/pages/public-sector/articles/blockchain-opportunities-for-health-care.html}
\BIBentrySTDinterwordspacing

\bibitem{Li2010}
M.~Li, S.~Yu, K.~Ren, and W.~Lou, ``{Securing personal health records in cloud
  computing: Patient-centric and fine-grained data access control in
  multi-owner settings},'' \emph{Lecture Notes of the Institute for Computer
  Sciences, Social-Informatics and Telecommunications Engineering}, vol. 50
  LNICST, pp. 89--106, 2010.

\bibitem{hu2013guide}
V.~C. Hu, D.~Ferraiolo, R.~Kuhn, A.~R. Friedman, A.~J. Lang, M.~M. Cogdell,
  A.~Schnitzer, K.~Sandlin, R.~Miller, K.~Scarfone \emph{et~al.}, ``Guide to
  attribute based access control (abac) definition and considerations
  (draft),'' \emph{NIST special publication}, vol. 800, no. 162, 2013.

\bibitem{godik2002oasis}
S.~Godik and T.~Moses, ``Oasis extensible access control markup language
  (xacml),'' \emph{OASIS Committee Secification cs-xacml-specification-1.0},
  2002.

\bibitem{Bogaerts:2015:EAC:2818000.2818009}
J.~Bogaerts, M.~Decat, B.~Lagaisse, and W.~Joosen, ``Entity-based access
  control: Supporting more expressive access control policies,'' in
  \emph{Proceedings of the 31st Annual Computer Security Applications
  Conference}, ser. ACSAC 2015.\hskip 1em plus 0.5em minus 0.4em\relax New
  York, NY, USA: ACM, 2015, pp. 291--300.

\bibitem{Di}
D.~Di~Francesco~Maesa, P.~Mori, and L.~Ricci, ``Blockchain based access
  control,'' \emph{Distributed Applications and Interoperable Systems: 17th
  IFIP WG 6.1 International Conference, DAIS 2017, Held as Part of the 12th
  International Federated Conference on Distributed Computing Techniques,
  DisCoTec 2017, Neuch{\^a}tel, Switzerland, June 19--22, 2017, Proceedings},
  pp. 206--220, 2017.

\bibitem{Yue2016}
X.~Yue, H.~Wang, D.~Jin, M.~Li, and W.~Jiang, ``{Healthcare Data Gateways:
  Found Healthcare Intelligence on Blockchain with Novel Privacy Risk
  Control},'' \emph{Journal of Medical Systems}, vol.~40, no.~10, 2016.

\bibitem{chepurnoyspace}
\BIBentryALTinterwordspacing
A.~Chepurnoy and D.~Meshkov, ``On space-scarce economy in blockchain systems,''
  \emph{{IACR} Cryptology ePrint Archive}, vol. 2017, p. 644, 2017. [Online].
  Available: \url{http://eprint.iacr.org/2017/644}
\BIBentrySTDinterwordspacing

\bibitem{nahitreport}
NAHIT, ``Report to the office of the national coordinator for health
  information technology on defining key health information technology terms,''
  Office of the National Coordinator for Health Information Technology, Tech.
  Rep., April 2008.

\end{thebibliography}

\end{document}